\documentclass[shortnote]{jpsj2} %% for short notes
\title{Calculation of the Self-energy of Open Quantum Systems}

\author{\textsc{Keita Sasada}$^{1}$\thanks{E-mail address: sasada@iis.u-tokyo.ac.jp} and \textsc{Naomichi Hatano}$^{2}$\thanks{E-mail address: hatano@iis.u-tokyo.ac.jp}}
\inst{$^{1}$Department of Physics, University of Tokyo, Komaba, Meguro, Tokyo 153-8505  \\
$^{2}$Institute of Industrial Science, University of Tokyo, Komaba, Meguro, Tokyo 153-8505 }
\kword{Open quantum system, Self-energy, Resonant state}
\begin{document}
\maketitle
%Introduction %%%%%%%%%%%%%%%%%%%%%%%%%%%%%%%%%%%%%%%%%%%%
The electronic conduction in mesoscopic systems has been studied extensively in recent years.
A theoretically interesting feature of the problem is the fact that the system in question is an open quantum system with semi-infinite leads.
The open quantum system intrinsically has resonant states, which can strongly affect the electronic conduction\cite{Sasada}. 

A popular way of treating the semi-infinite leads is to contract the leads to the self-energy.
The self-energy of leads is a useful way of computing the conductance as well as obtaining resonant states.
In this note, we propose a new method of calculating the self-energy of the leads.
The self-energy $\Sigma(E)$ was originally defined in\cite{Datta}
\begin{align}
\langle x|\frac{1}{E-H+i\delta}|x'\rangle=\langle x|\frac{1}{E-\left(H_{c}+\Sigma(E)\right)}|x'\rangle \label{eq:eff_Green}
\end{align}
for sites $x$ and $x'$ inside the central conductor, where $H_c$ is the Hamiltonian of the central conductor and $H$ is the total Hamiltonian including semi-infinite leads attached to the conductor.
The self-energy has been calculated by various methods. The method that we present here is much easier than previous methods.
The main claim of this note is that the self-energy is equivalent to the boundary conditions for resonant states.
%The self-enrgy of the one-dimensional leads%%%%%%%%%%%%%%%%%%%%%%%%%%%%%%%%%%%%%%%%%%%%%%%

We consider the Hamiltonian of a conductor with semi-infinite leads attached to it:
$H=H_c+\sum_{\alpha}H_{\alpha}, \label{eq:Hamiltonian}$
where $H_c$ is a one-body Hamiltonian of a finite-size conductor, while $H_{\alpha}$ describes a semi-infinite lead given by the tight-binding model
\begin{align}
H_{\alpha}\equiv\displaystyle{-t\sum_{x_{\alpha}=0}^{\infty}\left( |x_{\alpha}+1  \rangle\langle x_{\alpha}|+|x_{\alpha}\rangle\langle x_{\alpha}+1|\right)}.
\end{align}
This includes the hopping between a site $x_{\alpha}=0$ on the conductor and the lead $\alpha$. (Note that, if we have hopping between the conductor and a lead with the amplitude different from $-t$, we include it in $H_{c}$.)

Equation (\ref{eq:eff_Green}) suggests that the eigenvalues of the effective Hamiltonian $H_{\rm eff}(E)\equiv H_c+\Sigma(E)$ are the poles (bound states and resonant states) of the total Hamiltonian $H$ on the complex $E$ plane. 
Therefore, we seek discrete and generally complex eigenvalues $E_n$ of resonant states and bound states of the whole system:
\begin{align}
H |\psi_n\rangle=E_n|\psi_n\rangle \ \ \mbox{and} \ \ \ \langle\tilde{\psi}_n|H=E_n\langle \tilde{\psi}_n|.  \label{eq:Schrodinger}
\end{align}
The eigenfunctions are bi-orthogonal: $\langle \tilde{\psi}_n|\psi_m\rangle=\delta_{nm}$.
The eigenvalues $E_n$ are related to the corresponding eigen-wave-number $k_n$, which is also generally complex, through the dispersion relation $E_n=-2t \cos k_n$. 
The eigen-wave-number $k_n$ is on the upper-half plane for the bound states and on the lower-half plane for the resonant states.

It is known that the resonant states as well as the bound states can be found by requiring the boundary conditions $\langle x_{\alpha}|\psi_n\rangle \propto e^{ik_n x_{\alpha}}$ for $x_\alpha\geq 0$ in the leads\cite{Hatano}.
In other words, the discrete states satisfy the boundary conditions
\begin{align}
\langle x_{\alpha}+1|\psi_n\rangle=e^{ik_n}\langle  x_{\alpha}|\psi_n\rangle \ \ \ \mbox{for $x_{\alpha}\geq 0$},  \label{eq:boundary}
\end{align}
where $\Re k_n\geq 0$. 
The boundary conditions (\ref{eq:boundary}) transform the Schr\"{o}dinger equation
\begin{align}
\langle x_{\alpha}=0|H_c|\psi_n\rangle-t\langle x_{\alpha}=1|\psi_n\rangle= E_n\langle x_{\alpha}=0|\psi_n\rangle  \label{eq:Hamiltonian+space}
\end{align}
to
\begin{align}
\langle x_{\alpha}=0|H_c|\psi_n\rangle+V_{\rm eff}^{(\alpha)}(E_n)\langle x_{\alpha}&=0|\psi_n\rangle \nonumber\\
&= E_n\langle x_{\alpha}=0|\psi_n\rangle,\label{eq:Hamiltonian+boundary}
\end{align}
where
\begin{align}
V_{\rm eff}^{(\alpha)}(E)\equiv-te^{ik}\label{eq:eff_potential}
\end{align}
is the energy-dependent effective potential.

We claim that the self-energy of the lead $\alpha$ is nothing but the effective potential: 
\begin{align}
\Sigma^{(\alpha)}(E)=V_{\rm eff}^{(\alpha)}(E)|x_{\alpha}=0\rangle\langle x_{\alpha}=0|. \label{eq:one_self-energy}
\end{align}
The total self-energy is the sum over the leads: $\displaystyle{\Sigma(E)=\sum_{\alpha}\Sigma^{(\alpha)}(E)}$.
The effective potential $V_{\rm eff}^{(\alpha)}$ is rewritten in terms of $E$ as 
\begin{align}
V_{\rm eff}^{(\alpha)}(E)\equiv\frac{E-i\sqrt{4t^2-E^2}}{2}\label{eq:one_self-energy_E}
\end{align}
by using the dispersion relation $E=-2t\cos k$. Note that we choose the branch $\Im V_{\rm eff}^{(\alpha)}<0$ for the retarded Green function.
Equation (\ref{eq:one_self-energy_E}) is indeed equivalent to the expression obtained by other methods\cite{Datta}. 

\begin{figure}
\begin{center}
\includegraphics[width=0.4\textwidth]{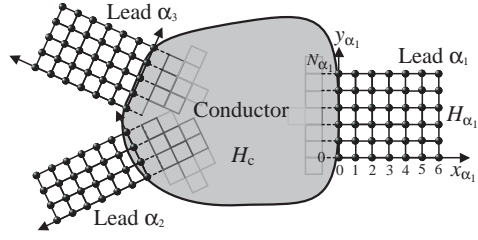}
\end{center}
\caption{Leads of the form of ladders are attached to the central conductor.}\label{fig:ladder}
\end{figure}

%The self-energy of N-legs ladder%%%%%%%%%%%%%%%%%%%%%%%%%%%%%%%%%%%%%%%%%%%%%%%%%
Let us now demonstrate that the present method is easily generalized to other types of leads such as {\it N}-leg ladder and carbon nanotube.
Hereafter, we drop the lead index $\alpha$ for simplicity.
First, we calculate the self-energy of a lead of {\it N}-leg ladder (Fig.\ref{fig:ladder}):
\begin{align}
H_{\rm ladder}=&-t\sum_{x=0}^{\infty}\sum_{y=1}^{N}\left( |x+1, y\rangle\langle x, y |\right. \nonumber \\
&\left. +|x, y+1\rangle\langle x, y |+{\rm c.c.}\right).
\end{align}
We first diagonalize $H_{\rm ladder}$ in the $y$ direction and obtain the conduction channels $\displaystyle{\left\{\phi_j\left(y\right)|j=1,2,\cdots,N\right\}}$,
where 
\begin{align}
\phi_{j}(y)=\sin\frac{j \pi y}{N+1}\Big/\sqrt{\sum_{y''=1}^{N}\sin^2 \frac{j \pi y''}{N+1}}. \label{eq:ladder_psi}
\end{align}
Each channel has the dispersion relation $E=-2t\cos k_j+\omega_j$, where $\omega_j\equiv-2t \cos(2\pi j/N)$.
Each channel yields its effective potential of the form Eq.~(\ref{eq:eff_potential}), or
\begin{align}
V_{\rm eff}^{(j)}(E)&=-te^{ik_j}=\frac{E-\omega_j-i \sqrt{4t^2-\left(E-\omega_j\right)^2}}{2}.
\end{align}
The self-energy of {\it N}-leg ladder is given in the $N\times N$ matrix form 
\begin{align}
\left(\Sigma_{\rm ladder}(E)\right)_{y,y'}&=\sum_{j=1}^{N}\phi_{j}(y)V_{\rm eff}^{(j)}(E)\phi_{j}(y')^{*}.
\end{align}
The result is equivalent to the one obtained in Ref.~\citen{Ladder}.
%The self-energy of zigzag carbon nanotube%%%%%%%%%%%%%%%%%%%%%%%%%%%%%%%%%%%%%%%%%%%%%%

\begin{figure}
\begin{center}
\includegraphics[width=0.32\textwidth]{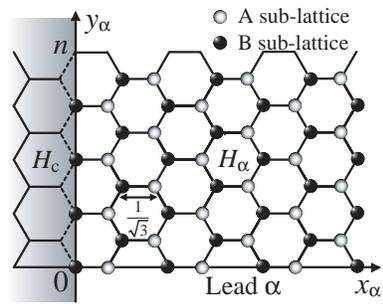}
\end{center}
\caption{A lead of the zigzag carbon nanotube. The upper and lower edges satisfy the periodic boundary conditions.}\label{fig:CNT_zigzag}
\end{figure}

Second, we calculate the self-energy of a lead of ({\it n},0) zigzag carbon nanotube attached to the conductor as in Fig.~\ref{fig:CNT_zigzag}, where $n$ is the chiral number. The Schr\"{o}dinger equation of the zigzag carbon nanotube
$H_{\rm zigzag}| \psi^{\pm}_{\rm A/B}(k_j)\rangle=E| \psi^{\pm}_{\rm A/B}(k_j)\rangle$ yields the dispersion relation of the {\it j}th channel as
\begin{align}
E&\displaystyle{=\pm t\left|h_{k_j}\right|}=\displaystyle{\pm t\sqrt{1\pm 4\cos\frac{\sqrt{3}k_{j}}{2}\cos\frac{\pi j}{n}+4\cos^{2}\frac{\pi j}{n}}},\label{eq:zigzag_dispersion}
\end{align}
with
\begin{align}
h_{k_j}\equiv e^{i\frac{k_j}{\sqrt{3}}}+2 \cos\frac{\pi j}{n} e^{-i\frac{k_j}{2\sqrt{3}}}
\end{align}
where the first Brillouen zone is $|k_{j}|<\pi/\sqrt{3}$~\cite{Riichiro},
and its wavefunction on the  A and B sub-lattices as
\begin{align}
\left\{
\begin{array}{ll}
\displaystyle{\langle x,y|\psi^{\pm}_{\rm A}(k_j)\rangle}=&\displaystyle{\mp \frac{h_{k_j}^{*}}{|h_{k_j}|}e^{i k_j x}\phi_j(y)},\\
\displaystyle{\langle x,y|\psi^{\pm}_{\rm B}(k_j)\rangle}=&\displaystyle{e^{i k_j x}\phi_j(y)} ,
\end{array}\label{eq:psi_zigzag}
\right.
\end{align}
where $\displaystyle{\phi_j(y)\equiv e^{i\frac{2\pi j}{n}y}\!/\sqrt{n}}$.
The boundary conditions (\ref{eq:psi_zigzag}) transform the Schr\"{o}dinger equation of the whole system
\begin{align}
\langle x=0,y|H_c|\psi_{\rm B}^{\pm}(k_j)\rangle-t\langle x&=1\big/\sqrt{3},y|\psi_{\rm A}^{\pm}(k_j)\rangle \nonumber \\
&=E\langle x=0,y|\psi_{\rm B}^{\pm}(k_j)\rangle \label{eq:Hamiltonian+space_zigzag}
\end{align}
to
\begin{align}
\langle x=0,y|H_c|\psi_{\rm B}^{\pm}(k_j)\rangle+V_{\rm zigzag}^{(j;B)}&(E)\langle x=0,y|\psi_{\rm B}^{\pm}(k_j)\rangle \nonumber \\
=& E\langle x=0,y|\psi_{\rm B}^{\pm}(k_j)\rangle , \label{eq:Hamiltonian+boundary_zigzag}
\end{align}
where the effective potential of the {\it j}th channel is given by
\begin{align}
&V_{\rm zigzag}^{(j;B)}(E)\equiv\pm t\frac{h_{k_j}^{*}}{|h_{k_j}|}e^{i \frac{k_j}{\sqrt{3}}} \\
&=\frac{E^2+t^2-\lambda_{j}^2\pm i\sqrt{\left(2 t \lambda_{j}\right)^2-\left(E^2-t^2-\lambda_{j}^2\right)^2}}{2 E}
\end{align}
with $\lambda_{j}\equiv 2 t \cos \pi j/n$.
Hence we obtain the self-energy of an ({\it n},0) carbon nanotube in the $n\times n$ matrix form
\begin{align}
\left(\Sigma_{\rm zigzag}(E)\right)_{y_{\rm B},y'_{\rm B}}&=\sum_{j=1}^{n}\phi_{j}(y_{\rm B})V_{\rm zigzag}^{(j;B)}(E)\phi_{j}(y'_{\rm B})^{*},\label{eq:self-energy_zigzag}
\end{align} 
where $y_{\rm A}$ and $y_{\rm B}$ are coordinates on the A and B sub-lattices, respectively, which are indicated in Fig.~\ref{fig:CNT_zigzag}.
The result (\ref{eq:self-energy_zigzag}) is indeed equivalent to the one obtained in Ref.~\citen{Zigzag}.

When the {\rm A} sub-lattice, instead of the {\rm B} sub-lattice, is in contact with the conductor,
we obtain the self-energy in the form
\begin{align}
\left(\Sigma_{\rm zigzag}(E)\right)_{y_{\rm A},y'_{\rm A}}&=\sum_{j=1}^{n}\phi_{j}(y_{\rm A})V_{\rm zigzag}^{(j;A)}(E)\phi_{j}(y'_{\rm A})^{*}\label{eq:self-energy_zigzag_A}
\end{align}
with
\begin{align}
&V_{\rm zigzag}^{(j;A)}(E)\nonumber \\
\equiv&\frac{E^2-t^2+\lambda_{j}^2\pm i\sqrt{\left(2 t \lambda_{j}\right)^2-\left(E^2-t^2-\lambda_{j}^2\right)^2}}{2 E}.
\end{align}

The authors are grateful to Dr. Manabu Machida for his helpful comments.
This work is supported by Grant-in-Aid for Scientific Research (No.17340115) from the Ministry of Education, Culture, Sports, Science  
and Technology as well as by Core Research for Evolutional Science and Technology (CREST) of Japan Science and Technology Agency.
%%%%%%%%%%%%%%%%%%%%%%%%%%%%%%%%%%%%%%%%%%%%%%

\end{document}